\documentstyle[preprint,eqsecnum,aps]{revtex}

\newcommand{\ubar}{\overline u}
\newcommand{\dbar}{\overline d}

\newcommand{\be}{\begin{equation}}
\newcommand{\ee}{\end{equation}}
\newcommand{\bear}{\begin{eqnarray}}
\newcommand{\ear}{\end{eqnarray}}
\newcommand{\la}{\label}

\input{epsf.sty}
\def\framegraphics{\def\ifframe{\iftrue}}
\def\dontframegraphics{\def\ifframe{\iffalse}}
\def\drawgraphics{\def\ifdraw{\iftrue}}
\def\dontdrawgraphics{\def\ifdraw{\iffalse}}

\dontframegraphics
\drawgraphics

\begin{document}
\draft
\title{
A Constraint on the $x$ Dependence of the Light Antiquarks Ratio
}
\author{F. M. Steffens\footnote{E-mail: fsteffen@if.usp.br},}
\address{Ericsson Telecommunications\\
Rua Maria Prestes Maia 300, CEP:02047-901 S{\~ a}o Paulo SP, Brazil
}
\date{\today}
\maketitle
\begin{abstract}

We perform a careful study on the effect of the Pauli blocking to the 
light antiquark structure of the proton sea. We develop the formal
expressions for the antiquark distributions, highlighting the  
role played by quark statistics and the vacuum structure. Ratios involving
the antiquarks are calculated. In particular, it is found that  
$\Delta\overline{d}(x)/\Delta\overline{u}(x)$ should be negative and $x$ independent.

\end{abstract}

\narrowtext

\hspace{1cm}

The first suggestion that the proton sea is not symmetric was put forward by 
Feynman and Field in 1977 \cite{field}. They noticed that because there are more
empty states for $d$ quarks in the proton, $\dbar d$ pairs would be created more easily 
than the $\ubar u$ pairs - the Pauli Blocking. In the early 1980's, 
Thomas \cite{thomas83}
noticed that a pion cloud in the proton having more $\pi^+$ than $\pi^-$ naturally
produces an excess of $\overline d$ antiquarks. Then in the early 1990's the first
measurements of the Gottfried sum rule by the NMC \cite{nmc} strongly suggested,
in the context of the Parton model, that there was indeed an excess of $\overline d$
over $\overline u$ antiquarks. A flurry of papers followed, where most of the work
was based on meson clouds (for a complete review on the Meson Cloud Model approach
to the light antiquark asymmetry see \cite{spethtony}), with a few exceptions where the Pauli
Blocking was advocated \cite{wally98,signal89}. Chiral quark models \cite{chiral} and 
instantons \cite{dorokhov} have also been used to explain the NMC data.
By now, extensive experimental data is available 
and we have at our disposition not only data on the magnitude of 
$\overline d - \overline u$, as well as on its $x$ dependence \cite{e866}.

In this letter we advocate the fundamental role played by the quark statiscs in both 
the polarized and unpolarized antiquark distributions.
In order to isolate such effects to the sea structure of the proton, we 
work out the formal definition of the antiquark distributions to a point where some 
actual predictions can be made without the introduction of a number of model parameters.
From this point, we are able to study to what extent Pauli Blocking is a correction
or is the main phenomenon behind the sea assimmetry of the light quarks. 

We start with the formal definition of the antiquark distribution \cite{jaffe83}:

\be
\overline q(x) = \frac{p^+}{2\pi}\int dz^- e^{-ixp^+ z^-}
<P|\psi_+ (z^-) \psi^\dagger_+ (0) |P>_c{\huge\mid}_{z^+ = z^\perp = 0},
\la{e1}
\ee
with $x > 0$. To calculate Eq. (\ref{e1}), we remember that the sum over all 
the possible spins of the $\psi_+$ operators provides:

\be
\psi_+^\dagger \psi_+ = \psi_+^{\dagger\uparrow} \psi_+^\uparrow +
\psi_+^{\dagger\downarrow} \psi_+^\downarrow = 
\frac{1}{2} \psi^\dagger (1 + \gamma_0 \gamma_3)\psi,
\la{e3}
\ee
with

\be
\psi (z) = \int \frac{d^3k}{(2\pi)^3} 
\sum_\alpha [b_\alpha (k) u^{(\alpha)} (k) e^{-ik\cdot z} + 
d_\alpha^\dagger (k) v^{(\alpha)} (k) e^{ik\cdot z}].
\la{e4}
\ee
Using Eq. (\ref{e4}) in Eq. (\ref{e1}) we see that the term which
involves the quark operators produces a restriction, in the form 
of $\delta (xp^+ + k^+)$, which forces it to not contribute to the antiquark
distributions, once $x$ should be positive, ($p^+$ and $k^+$
are positive). Hence, taking a general for the Dirac spinor, with
$f(k)$ and $g(k)$ its upper and lower components, respectively, 
the antiquark distribution is rewritten as\footnote{As usual in the definition of parton distributions, 
we take the parton transverse momenta squared not to be large. Or, correspondingly, 
we take $k_z^2 >> k_{\perp}^2$, implying that $\sigma\cdot \vec k \chi_\alpha = 
\pm \chi_\alpha$ (+ for $\alpha = \uparrow$; - for $\alpha = \downarrow$)}:

\be
\overline q(x) = \frac{1}{2}\int \frac{d^3 k}{(2\pi)^3}
[f^2 + g^2 + 2f\cdot g]
<p|d^\dagger_\uparrow (k) d_\uparrow (k) +
d^\dagger_\downarrow (k) d_\downarrow (k) |p>_c 
\delta (x - \frac{k^+}{p^+}).
\la{e10}
\ee
The calculation of the polarized distributions follows the same pattern.
From the expression for the polarized antiquark distribution:

\be
\overline q(x) = \frac{p^+}{2\pi}\int dz^- e^{-ixp^+ z^-}
<P|\psi_+ (z^-)\gamma_5 \psi^\dagger_+ (0) |P>_c{\huge\mid}_{z^+ = z^\perp = 0},
\la{e18}
\ee
for $x>0$, we get:

\be
\Delta\overline q(x) = \frac{1}{2}\int \frac{d^3 k}{(2\pi)^3}
[f^2 + g^2 + 2f\cdot g]
<p|d^\dagger_\uparrow (k) d_\uparrow (k) -
d^\dagger_\downarrow (k) d_\downarrow (k) |p>_c 
\delta (x - \frac{k^+}{p^+}).
\la{e21}
\ee

We expanded the $\psi$ operators in terms of free quark and antiquark operators.
Quarks in the proton, however, are not free and that can be translated into 
a complicated vacuum structure where the confined quarks 
live \cite{chiral,kazuo01,diakonov}.
In this environment, the free space vacuum is certainly not a good 
approximation for the ground state of the confined quark operators.
Instead, one has to build a new vacuum structure based on bound
quark operators.
Following this direction, we extend the work of Tsushima, Thomas and Dunne 
\cite{kazuo01},
to derive general forms for the antiquark distributions 
which incorporate the effects from the modified vacuum.
The bound state operators (denoted by a ``*'') and the free state operators
are related by a Bogolyubov transformation:

\bear
b^* &=& A\cdot b + B\cdot d^\dagger ,\\ \nonumber
d^{*\dagger} &=& C \cdot b + D\cdot d^\dagger ,
\la{e22}
\ear
where the $A$, $B$, $C$, $D$ factors are the overlaps
between the bound and free states. As shown in \cite{kazuo01},
these overlaps are non zero for a confining scalar potential,
at least in 1+1 dimensions. 
The vacuum of the bound states are defined such that:

\bear
b^* |0^*> &=& 0 , \\ \nonumber
d^* |0^*> &=& 0.
\la{e23}
\ear
From Eqs. (\ref{e22}) and (\ref{e23}) it follows that:

\bear
b|0^*> &=& - \frac{B}{DA - CB}|\overline{q}^*>, \\ \nonumber
d|0^*> &=& - \frac{C^\dagger}{D^\dagger A^\dagger - C^\dagger B^\dagger}|q^*>,
\la{e24}
\ear
where $|\overline{q}^*>=d^{*\dagger}|0^*>$, and $|q^*>=b^{*\dagger}|0^*>$.
Eqs. (\ref{e24}) are telling us that the modified vacuum is not empty but
filled with quark-antiquark pairs.

In order to reach any conclusion regarding the sea quark structure
of the proton, we should specify what is the state $|p>$ in 
Eqs. (\ref{e10}) and (\ref{e21}). The proton state, as a bound state
of quarks, has to be built from bound state quark operators acting 
on the modified vacuum, as described by Eq. (\ref{e23}). The simplest form one
can use for such state is that given by the $SU(6)$ wave function:

\bear
|p\rangle &\equiv& {\it F[b^{*\dagger}]}|0^*\rangle \\ \nonumber
&=& \frac{1}{\sqrt{18}}\epsilon^{\alpha \beta \gamma}[b^{*\dagger} (u,\uparrow,\alpha)
b^{*\dagger} (d,\downarrow,\beta) - b^{*\dagger} (u,\downarrow,\alpha)
b^{*\dagger} (d,\uparrow,\beta)] b^{*\dagger} (u, \uparrow, \gamma) |0^*\rangle .
\la{e11}
\ear
Because we know, from Eqs. (\ref{e24}) how the free quark operators act on the 
modified vacuum, we can readily calculate the distributions written in Eqs.
(\ref{e10}) and (\ref{e21}). In particular, our interest is the antiquark distributions,
which is calculated to be:

\be
\overline{q}^m(x) = \frac{1}{2}\int \frac{d^3 k}{(2\pi)^3}
[f^2 + g^2 + 2f\cdot g]\frac{|C|^2}{|BC - AD|^2}
<q^*_m|{\it F^\dagger [b^{*\dagger}]} {\it F[b^{*\dagger}]}|q^*_m>\delta (x - \frac{k^+}{p^+}),
\la{e111}
\ee
where $m$ stands for the possible spin projections.
Notice that the antiquark distribution comes from the expectation values between 
the bound quark states. In the case of a proton state built from free quarks, there is
no antiquark distributions, as $\frac{|C|^2}{|BC - AD|^2}\rightarrow 0$. 
We can, of course, generate the sea
quarks through perturbation theory once we know the quark - gluon
vertex from QCD, a procedure which has actually been implemented  
before \cite{golowich,steffens96}. In this case, it happens that
the quark-antiquark pairs generated from perturbative gluons produce
an excess of $\overline u$ antiquarks over $\overline d$ antiquarks -
in clear contradiction to the naive expectation from the Pauli principle.
The solution to this dilemma is straightforward and will be presented in a forthcoming work.

To proceed, we need the calculation of the expectation value between the bound 
quark states. A direct calculation shows that:

\be
<u^*|{\it F^\dagger [b^{*\dagger}]} {\it F[b^{*\dagger}]}|u^*> = 4,
<d^*|{\it F^\dagger [b^{*\dagger}]} {\it F[b^{*\dagger}]}|d^*> = 5,
\la{e112}
\ee
where the sum over the spins have been performed. Thus it follows that there is an
excess of down antiquarks in the proton compared to the up antiquarks. 
Now we can write down the explicit expression for the difference:

\be
\overline d(x) - \overline u(x) = \frac{1}{2}\int \frac{d^3 k}{(2\pi)^3}
[f^2 + g^2 + 2f\cdot g]\frac{|C|^2}{|BC - AD|^2}\delta (x - \frac{k^+}{p^+}),
\la{e113}
\ee
which is positive definite. Although we do not know how to calculate 
Eq. (\ref{e113}) for the realistic (3 + 1) dimensions case, we can make 
a direct comparison to the experimental result by calculating the ratio:

\be
\frac{\overline d(x)}{\overline u(x)} = \frac{5}{4}.
\la{e1131}
\ee
The experimental result is not an $x$ independent curve. However,
we know that in the small $x$ region, where the number of perturbative
antiquarks is quite large, the ratio should approach 1. In the intermediate
$x$ region the number of perturbative antiquarks
diminishes drastically and the constant ratio, around the 5/4 of 
Eq. (\ref{e1131}), should appear. However, inside this region we have
the meson cloud, which is known to give its main contribution around the 
$x=0.2$ region. The deviation from 5/4 is maximum
exactly in this region. Nevertheless, the interesting point is that
the result encapsulated by Eq. (\ref{e1131}) is fundamental for any 
calculation of the sea asymmetry effect.

Having established the fundamental role played by the quark statistics,
along with the vacuum structure, to the observed sea asymmetry, we discuss
the role of the Pauli Blocking in the polarized sea.
As before, we calculate the expectation value for the number of polarized antiquarks
of a given flavor in the proton using the quark distribution as given in (\ref{e111}). 
However, instead of summing over the spins, we
now have to take the difference. The matrix elements in Eq. (\ref{e111}) 
when calculated for $m=\uparrow$ and $m=\downarrow$ gives a factor of
4/3 for the $\overline u$ antiquarks and 1/3 for the $\overline d$ antiquarks.
Hence, we have for the polarized distributions:

\bear
\Delta\overline u(x) &=& \frac{4}{3}\frac{1}{2}\int \frac{d^3 k}{(2\pi)^3}
[f^2 + g^2 + 2f\cdot g]\frac{|C|^2}{|BC - AD|^2}\delta (x - \frac{k^+}{p^+}), \\
\Delta\overline d(x) &=& -\frac{1}{3}\frac{1}{2}\int \frac{d^3 k}{(2\pi)^3}
[f^2 + g^2 + 2f\cdot g]\frac{|C|^2}{|BC - AD|^2}\delta (x - \frac{k^+}{p^+}).
\la{e1132}
\ear
Note that the ratios, like

\be
\frac{\Delta\overline d(x) - \Delta\overline u(x)}{\overline d(x) - \overline u(x)} = -\frac{5}{3},
\la{e114}
\ee
are $x$ independent up to pionic corrections to the unpolarized distributions and
quark mass corrections to all distributions. 
A different choice for the wave function, Eq. (\ref{e11}), would render a different numerical 
factor, but it would not introduce an extra $x$ dependence in the ratio. On the other hand, the ratio 

\be
\frac{\Delta\overline d(x)}{\Delta\overline u(x)} = -\frac{1}{4}
\la{e115}
\ee
should not be affected by pions (as they do not contribute to the polarized 
distributions). In this case, the experimental value for 
$\Delta\overline d(x)/ \Delta\overline u(x)$ should be very close to a straight line.

We have seen that developing the formal expression for the antiquark distributions,
Eq. (\ref{e1}), as far as possible in terms of the quark operators, allows us to 
derive some general conclusions about the physics responsible for the antiquark asymmetries
in the proton sea. We see that the vacuum structure inside the proton, as expressed by
Eqs. (\ref{e24}), is decisive in reaching this conclusion. However, it is not the 
final story, as Eqs. (\ref{e24}) would exist even if quarks were bosons. 
The fact that quarks are fermions is fundamental for the observed asymmetry in the
number of the unpolarized light quarks in the proton sea, as can be seen from 
Eqs. (\ref{e112}). For the polarized case, the quark statistics is more than fundamental, 
it is probably the only sizeable effect to be measured. 
Of course, to extract numbers we need to model the proton wave function.
Using the SU(6) quark wave function for the proton, we see that
the $x$ dependence cancels when ratios are taken.
The result embodied by Eq. (\ref{e115}) is made more important when we remember the discussion after
the $\overline d(x)/\overline u(x)$ was calculated: pion dressing of the proton wave function is the 
effect that gives the $x$ dependence of the unpolarized ratio. It means that the ratio would
be different from the unity even in a world without pions. In the 
polarized case, apart from quark mass effects in the large $x$ region,
Eq. (\ref{e115}) should hold. In this way, the polarized ratio is expected to be $x$ 
independent.
\newline
\newline
\newline
I would like to thank W. Melnitchouk and Kazuo Tsushima for helpful discussions.

\addcontentsline{toc}{chapter}{\protect\numberline{}{References}}

\end{document}